\documentclass[12pt]{article}

\usepackage[preprint, nonatbib]{neurips_2023}

\usepackage[utf8]{inputenc} 
\usepackage[T1]{fontenc}    
\usepackage{hyperref}       
\usepackage{url}            
\usepackage{booktabs}       
\usepackage{amsfonts, amsmath}       
\usepackage{nicefrac}       
\usepackage{microtype}      
\usepackage{xcolor}         

\usepackage{fancyhdr}

\usepackage{algorithm}
\usepackage{algpseudocode}

\hypersetup{colorlinks, linkcolor=blue, citecolor=blue, urlcolor=blue}

\usepackage{amsfonts,dsfont}
\usepackage{hyperref}
\usepackage{color}
\usepackage{amsmath, amssymb}
\usepackage{amsthm}
\usepackage{bm}
\usepackage{stmaryrd}
\usepackage{color}
\usepackage{comment}
\usepackage{geometry}
\allowdisplaybreaks

\usepackage{graphicx}
\usepackage{caption}
\usepackage{subcaption}

\usepackage{tabularx}

\usepackage{caption}

\newcommand{\todo}[1][\null]{\ensuremath{\clubsuit}}

\newcommand{\noprint}[1]{}

\theoremstyle{definition}
\theoremstyle{definition}
\theoremstyle{definition}
\theoremstyle{definition}
\theoremstyle{definition}
\theoremstyle{definition}
\theoremstyle{definition}
\theoremstyle{definition}
\theoremstyle{definition}

\usepackage{tikz}
\usetikzlibrary{positioning, arrows.meta, calc}

\title{A solver-in-the-loop framework for end-to-end differentiable coastal hydrodynamics}

\author{%
  Elsa Cardoso-Bihlo \\
  Department of Mathematics and Statistics\\
  Memorial University of Newfoundland\\
  St. John’s, NL, A1C 5S7, Canada \\
  \texttt{ecardosobihl@mun.ca} \\
  \And
  Alex Bihlo \\
  Department of Mathematics and Statistics\\
  Memorial University of Newfoundland\\
  St. John’s, NL, A1C 5S7, Canada \\
  \texttt{abihlo@mun.ca} \\
}

\begin{document}

\maketitle

\vspace{0.25cm}

\begin{abstract}
Numerical simulation of wave propagation and run-up is a cornerstone of coastal engineering and tsunami hazard assessment. However, applying these forward models to inverse problems, such as bathymetry estimation, source inversion, and structural optimization, remains notoriously difficult due to the rigidity and high computational cost of deriving discrete adjoints. In this paper, we introduce \texttt{AegirJAX}, a fully differentiable hydrodynamic solver based on the depth-integrated, non-hydrostatic shallow-water equations. By implementing the solver entirely within a reverse-mode automatic differentiation framework, \texttt{AegirJAX} treats the time-marching physics loop as a continuous computational graph. We demonstrate the framework's versatility across a suite of scientific machine learning tasks: (1) discovering regime-specific neural corrections for model misspecifications in highly dispersive wave propagation; (2) performing continuous topology optimization for breakwater design; (3) training recurrent neural networks in-the-loop for active wave cancellation; and (4) inverting hidden bathymetry and submarine landslide kinematics directly from downstream sensor data. The proposed differentiable paradigm fundamentally blurs the line between forward simulation and inverse optimization, offering a unified, end-to-end framework for coastal hydrodynamics.
\end{abstract}

\section{Introduction}

Accurate modeling of free-surface flow, wave propagation, and coastal run-up is critical for assessing tsunami flood hazards, designing coastal infrastructure, and understanding near-shore hydrodynamics. Modern computational coastal engineering relies heavily on numerical solvers derived from the nonlinear shallow-water and Boussinesq-type equations~\cite{marr20a}. While the forward discretization of these models has reached a high degree of maturity, applying them to inverse problems, such as inferring unobservable initial bathymetry, discovering missing sub-grid physics, or designing optimal topological structures, remains a substantial computational bottleneck.

Traditionally, gradient-based inverse problems in fluid mechanics are solved using the adjoint-state method~\cite{sand00a}. While efficient for computing sensitivities with respect to high-dimensional parameter spaces, deriving and implementing the discrete adjoint equations for a complex, non-linear, non-hydrostatic fluid solver is exceptionally labor-intensive and error-prone~\cite{alha25a}. This difficulty is severely compounded in coastal hydrodynamics. Implicit pressure solvers (such as the Poisson equation required for non-hydrostatic pressure components) often rely on iterative matrix preconditioning that breaks continuous adjoint derivations. Furthermore, standard numerical treatments of discontinuous phenomena, such as the boolean flags used to track moving wet--dry shorelines during inundation, sever the flow of gradients entirely, rendering gradient-based optimization near the coast mathematically ill-posed. While continuous and recursive adjoint models have been developed for higher-order spectral wave simulations to reconstruct bathymetry~\cite{wu23a}, deriving these terms requires expanding the water wave equations to specific perturbation orders, rendering them computationally inflexible when boundary conditions or friction parameterizations change.

To bypass the rigidity of traditional discrete adjoints, the field of computational mechanics has increasingly embraced scientific machine learning and differentiable programming~\cite{sapi24a}. Recent advancements have popularized physics-informed neural networks and neural operators for fluid dynamics~\cite{anel21a,bihl22a,brec25a,kepp25a,leit21a,qi24a}. While these purely data-driven or soft-constrained models show promise, they frequently struggle to resolve multi-scale coastal phenomena, such as highly dispersive super-harmonics or discontinuous moving waterlines, due to spectral bias and the difficulty of balancing competing loss terms within the optimization landscape~\cite{wang22b}. In contrast, the \textit{solver-in-the-loop} paradigm~\cite{koch21a,um20a} adopted in this work retains the rigorous, momentum-conserving discretization of classical numerical solvers. By writing the numerical solver entirely in a framework supporting reverse-mode automatic differentiation~\cite{bayd18a}, the solver itself functions as a heavily structured, physics-informed computational graph. Exact analytical gradients can be back-propagated through the unrolled temporal simulation directly to the inputs, boundary conditions, or parameterized physical models.

Recent works in computational fluid dynamics have demonstrated that training neural networks in the loop produces localized corrections that are stable and respect the underlying numerical truncation errors, enabling accurate simulations on heavily coarsened grids~\cite{koch21a,koch24a}. Furthermore, the broader application of automatic differentiation to inverse design has recently gained traction in complex fluid-structure interactions~\cite{alha25a}, demonstrating that end-to-end differentiable simulators can solve high-dimensional optimization problems at a fraction of the cost of traditional adjoint methods.

This differentiable paradigm also unlocks advanced capabilities in active control and structural design. While topology optimization using the Solid Isotropic Material with Penalization (SIMP) method is mature in solid mechanics~\cite{bend13a}, its application in coastal engineering is mostly limited to heuristic methods or simplified linear wave equations. By utilizing a fully differentiable nonlinear fluid solver, continuous topology optimization for coastal structures can be performed directly against transient wave forcing. Similarly, for the active control of coastal structures, traditional approaches either rely on simplified linear control theory or utilize model-free reinforcement learning. In recent years, deep reinforcement learning has been increasingly explored for the active control of wave energy converters and autonomous wave-makers~\cite{ande20a,faed17a}. Model-free reinforcement learning, however, is notoriously sample-inefficient because it must learn the underlying fluid physics entirely through trial and error~\cite{holl20a}. Integrating a differentiable partial differential equations solver into the training loop allows a neural policy to interact with the exact analytical gradients of the fluid dynamics, enabling the efficient learning of robust control policies over long time frames.

In this paper, we present \texttt{AegirJAX}, a fully differentiable depth-integrated, non-hydrostatic solver based on the work of~\cite{stel03a,yama09a} designed specifically to bridge the gap between complex coastal mechanics and modern machine learning optimization. The organization of this paper is as follows. In Section~\ref{sec:LiteratureReview}, we provide a summary of the relevant literature on differentiable solver paradigms and tsunami propagation models. Section~\ref{sec:Theory} details the governing physical equations, the time-splitting predictor-corrector discretization, and the specific algorithmic modifications required to preserve end-to-end differentiability. In Section~\ref{sec:NumericalExperiments}, we move beyond the standard forward/inverse dichotomy to demonstrate the capabilities of \texttt{AegirJAX} across four distinct scientific machine learning domains: model discovery and correction, topological inverse design, active policy-in-the-loop control, and source parameter estimation. Finally, Section~\ref{sec:Conclusion} summarizes our findings and discusses the current limitations and future trajectories of data-driven physical correction.

\section{Related work}\label{sec:LiteratureReview}

An extensive array of numerical solvers for models of varying levels of complexity have been proposed for wave propagation and run-up modeling over the past several decades. Models can be broadly classifying by the types of governing equation they are solving, their employed turbulence models, their capabilities of handling breaking waves, and their spatial and temporal discretization schemes being used, see~\cite{marr20a} for a recent review. A wide variety of finite-difference methods~\cite{goto97a,lyne02a,tito97Ay,tito98a}, finite elements and discontinuous Galerkin methods~\cite{bokh05a,bone18a,lai12a,vate15a}, meshless methods~\cite{bihl17a,brec18a} and smoothed particle hydrodynamics based approaches~\cite{stge14a} have been used for this purpose. While traditional numerical discretization methods dominate wave propagation and inundation modeling, recent years have seen an increasing number of models being proposed that rely on neural-network based approaches, either using purely data-driven methodologies~\cite{liu21a,muli20a,muli22a} or physics-informed approaches~\cite{anel21a,bihl22a,brec25a,kepp25a,leit21a,qi24a,some25a}.

The estimation of unknown parameters in coastal engineering, such as initial sea-surface deformations, bottom bathymetry, or basal friction, has traditionally been framed as an optimization problem governed by shallow-water or Boussinesq-type partial differential equations. Gradient-based optimization using the adjoint-state method has been the undisputed method of choice for large-scale hydrodynamic inversions~\cite{navo98a}. Adjoint models allow for the efficient computation of gradients of an objective function with respect to distributed model parameters, requiring only one forward and one backward simulation regardless of the number of parameters. This methodology has been successfully applied to tsunami source inversion, allowing for the direct optimization of fault parameters from tide-gauge data~\cite{hoss18a,pire01a}. It has also been widely utilized for bathymetry estimation from surface elevation and velocity observations using the two-dimensional shallow-water equations~\cite{monn16a}. However, deriving the continuous or discrete adjoint equations for non-linear, dispersive wave models is a complex, mathematically challenging, and computationally expensive~\cite{alha25a}. Furthermore, standard numerical treatments of discontinuities, such as boolean wet--dry masks used for inundation, often sever the flow of gradients entirely~\cite{gile03a,monn16a}.

To bypass the rigidity of traditional discrete adjoints, the field of scientific machine learning has embraced differentiable programming~\cite{sapi24a}. By implementing established numerical discretization schemes within frameworks that support reverse-mode automatic differentiation, the entire time-marching physics loop becomes a continuous computational graph. This \textit{solver-in-the-loop} approach~\cite{koch21a,um20a} facilitates the framework of universal differential equations, where known physical laws (like mass and momentum conservation) are retained explicitly as discretized partial differential equations, while unknown, unresolved, or sub-grid physical processes are parameterized by neural networks embedded directly within the numerical operations~\cite{rack20a}. Recent works in computational fluid dynamics have demonstrated that training neural networks in the loop with a differentiable solver produces localized corrections that are stable and respect the underlying numerical truncation errors, enabling accurate simulations on heavily coarsened grids~\cite{koch21a,koch24a,whit25a}. Furthermore, the broader application of automatic differentiation to inverse design has recently gained traction in complex particle-laden flows and fluid-structure interactions, demonstrating that end-to-end differentiable simulators can solve high-dimensional optimization problems at a fraction of the cost of traditional adjoint methods~\cite{alha25a}. Differentiable solvers also offer profound advantages for the active control of fluid systems. While traditional model-free reinforcement learning must learn the underlying fluid physics entirely through trial and error, integrating a differentiable partial differential equations solver into the training loop allows a neural policy to interact with the exact dynamics during training~\cite{holl20a}. This approach allows for the efficient learning of control policies for complex nonlinear physical systems over long time frames. Controllers trained in this manner reliably generalize to unseen forcing conditions and can achieve stable rollouts over hundreds of recurrent evaluation steps, significantly outperforming derivative-free optimization.

\section{Governing equations and fully differentiable numerical discretization}~\label{sec:Theory}

In this section we described the governing equations that we employ for the subsequent study, as well as some general information about the differentiability paradigm employed, and what changes it necessitates to the discretization of the governing equations used.

\subsection{Governing equations}

The core hydrodynamic engine of \texttt{AegirJAX} is based on the depth-integrated non-hydrostatic shallow-water equations originally proposed by Stelling and Zijlema~\cite{stel03a} and later extended by Yamazaki et al.~\cite{yama09a} into the NEOWAVE model. \texttt{AegirJAX} can be seen as a differentiable extension to the NEOWAVE model, although we stress that the differentiable paradigm proposed here is independent of the particular physical model and numerical discretization being utilized. 

The governing equations are derived from the incompressible Navier--Stokes equations by assuming a linear vertical velocity profile and decomposing the total pressure into hydrostatic and non-hydrostatic components. The resulting continuity and horizontal momentum equations solved in \texttt{AegirJAX} are
\begin{subequations}\label{eq:GoverningEquations}
\begin{align}
\begin{split}
&\frac{\partial \zeta}{\partial t} + \nabla \cdot (\mathbf{U}D) = 0,\\
&\frac{\partial \mathbf{U}}{\partial t} + (\mathbf{U} \cdot \nabla)\mathbf{U} = -g\nabla\zeta - \frac{1}{2\rho}\nabla q - \frac{q}{2\rho D}\nabla(\zeta - h) - \mathbf{F}_f,
\end{split}
\end{align}
where $\zeta$ is the free-surface elevation, $h$ is the still water depth, $D=\zeta+h$ is the total flow depth, $U=(U,V)$ is the depth-averaged horizontal velocity vector, $\rho$ is the water density, $q$ is the non-hydrostatic pressure component, and $g$ is the gravitational acceleration. The term $\mathbf{F}_f$ represents an arbitrary forcing vector. While traditionally $\mathbf{F}_f$ is used strictly to model bottom friction via Manning's roughness formulation, within this differentiable framework it is explicitly generalized. It accommodates both classical empirical friction and dynamically parameterized forcing, such as localized momentum corrections injected by a neural network to account for unresolved, sub-grid physics. The non-hydrostatic pressure $q$ at the bottom is evaluated implicitly through the depth-integrated vertical momentum equation and the divergence-free condition
\begin{align}
\frac{\partial W}{\partial t} = \frac{q}{\rho D},\quad
\nabla \cdot \mathbf{U} + \frac{w_{\rm s} - w_{\rm b}}{D} = 0,
\end{align}
\end{subequations}
where $W$ is the depth-averaged vertical velocity, and $w_{\rm s}$ and $w_{\rm b}$ are the vertical velocities at the free surface and bottom, respectively, defined by the kinematic boundary conditions.

\subsection{The differentiable solver paradigm}

Traditional approaches to hydrodynamic optimization and inverse problems generally rely on adjoint methods~\cite{navo98a,sand00a}. While powerful, deriving and implementing the discrete adjoint equations for a complex, non-linear system like the Boussinesq equations is notoriously labor-intensive, rigid, and error-prone. Any change to the forward numerical scheme, boundary conditions, or friction parameterization requires a re-derivation of the corresponding adjoint solver.

Differentiable programming offers a paradigm shift by treating the numerical solver itself as a structured, physics-informed neural network layer. Rather than deriving analytical adjoints, a differentiable solver utilizes \textit{reverse-mode automatic differentiation}~\cite{bayd18a}. As the forward simulation executes, the underlying framework records the sequence of elementary mathematical operations into a computational graph. During the backward pass, the chain rule is applied systematically through this graph to compute the exact gradients of a user-defined scalar objective function (e.g., the difference between simulated and observed wave gauges) with respect to any input parameter, initial condition, or boundary state.

This approach yields several critical advantages for coastal hydrodynamic modeling:

\begin{description}
    \item[Exact gradients:] Unlike finite-difference approximations, automatic differentiation provides analytical-level precision for gradients (modulo roundoff errors due to finite-precision arithmetic), avoiding truncation errors and scaling to millions of parameters.
    \item[Arbitrary objectives:] The loss function can be modified on the fly (e.g., adding Total Variation regularization to a bathymetry inversion or adding an energy penalty to an active control task) without altering the underlying gradient calculation engine.
    \item[Seamless machine learning integration:] Because the solver is written in the same automatic differentiation framework used for deep learning, standard neural network architectures (e.g., convolutional layers or recurrent cells) can be coupled directly to the fluid state. The gradients flow unbroken from the downstream hydrodynamic loss, through the fluid domain, and directly into the neural network's weights.
\end{description}

However, constructing a solver that fully exploits this paradigm requires careful consideration of the computational graph. Traditional numerical treatments of discontinuities, such as abrupt wet--dry boundary flags or iterative matrix preconditioning, can sever the flow of gradients. To preserve end-to-end differentiability, the numerical discretization must be reformulated accordingly.

\subsection{Numerical discretization and implementation}

System~\eqref{eq:GoverningEquations} is solved using a time-splitting predictor--corrector scheme, following the methodology employed by NEOWAVE~\cite{yama09a}. We discretize the governing equations on a space-staggered Arakawa C-grid, where the scalar variables ($\zeta, h, q, w_{\rm s}, w_{\rm b}$) are located at the cell centers, and the horizontal velocities ($U, V$) are defined at the cell interfaces. The temporal discretization utilizes a semi-implicit fractional-step method consisting of a hydrostatic predictor and a non-hydrostatic corrector.

The intermediate horizontal velocities, $\tilde{\mathbf{U}}^{m+1}$, are computed explicitly by neglecting the non-hydrostatic pressure gradient terms in the momentum equations. To accurately capture flow discontinuities associated with breaking waves and hydraulic jumps without artificial dissipation, we utilize the momentum-conserved advection (MCA) scheme proposed in~\cite{stel03a}. For the continuity equation, we employ the upwind flux approximation, which determines the advective fluxes by extrapolating the free-surface elevation $\zeta$ rather than the total flow depth $D$. This significantly improves the numerical stability in the presence of strong discontinuous flows.

To account for wave dispersion, the intermediate velocity field is corrected using the non-hydrostatic pressure $q^{m+1}$. By substituting the discrete momentum updates into the continuity equation, a Poisson-type equation for the non-hydrostatic pressure is formulated,
\[
P q^{m+1} = Q(\tilde{\mathbf{U}}^{m+1}, w_{\rm s}^m, w_{\rm b}^{m+1}),
\]
where $P$ is an asymmetric penta-diagonal matrix (in 2D) depending strictly on the geometry of the previous time step, and $Q$ is the forcing vector formed by the divergence of the predicted hydrostatic velocity field. Once $q^{m+1}$ is obtained, the horizontal and vertical velocities are corrected, and the free-surface elevation $\zeta^{m+1}$ is updated.

While the mathematical foundation of \texttt{AegirJAX} mirrors classical non-hydrostatic models, the requirement for reverse-mode automatic differentiation necessitates several fundamental departures from traditional discrete implementations. To ensure that gradients can flow through the entire time-marching loop, the solver was written purely in \texttt{JAX}~\cite{brad18a}, replacing non-differentiable control flow and discrete state-tracking with continuous, array-based operations.

\medskip
\noindent\textbf{Differentiable wet--dry interface condition.} Many traditional inundation models handle moving waterlines using discrete boolean flags to track wet and dry cells, activating or deactivating fluxes based on threshold depths. While the underlying physical predictor--corrector steps in \texttt{AegirJAX} retain this discrete treatment to maintain standard numerical stability, these hard branching operations yield zero or undefined gradients. This effectively severs the backpropagation necessary for optimizing data-driven models at the shoreline. To resolve this, \texttt{AegirJAX} introduces an auxiliary continuously differentiable wet--dry masking function specifically designed to modulate the injected neural momentum corrections. Instead of a discrete switch, the wet fraction of a computational cell is approximated using a scaled sigmoid function
\[
M_{j,k} = \sigma(S \cdot (D_{j,k} - h_{\rm min})),
\]
where $h_{\rm min}$ is the minimum depth threshold and $S$ is a steepness parameter controlling the sharpness of the transition. When mapping cell-centered neural accelerations to the staggered velocity interfaces, this continuous mask is applied multiplicatively (e.g., $M_{i,j} \cdot M_{i+1,j}$), functioning as a soft logical \texttt{AND} operator. Consequently, the corrective forcing gracefully tapers to zero at the moving waterline without breaking the computational graph, allowing the solver to smoothly compute gradients with respect to inundation metrics (e.g., maximizing run-up or minimizing harbor impact) as illustrated in Section~\ref{sec:NumericalExperiments}.

\medskip
\noindent\textbf{Neural forcing integration.} To facilitate the model discovery tasks discussed in Section~\ref{sec:NumericalExperiments}, the solver architecture includes a dedicated fractional step for integrating data-driven momentum corrections. Rather than modifying the core analytical equations, the neural network acts as a corrective forcing applied after the physical predictor-corrector steps. This is realized via the computation of a \textit{neural network based} acceleration field, denoted by $\mathbf{a}_{nn}$, that aims to correct any model mispecifications not captured in the predicted hydrostatic and non-hydrostatic accelerations.

Let $\tilde{U}^{m+1} = (\tilde{U}^{m+1}, \tilde{V}^{m+1})$ denote the intermediate velocity field obtained after the hydrostatic predictor and non-hydrostatic pressure corrector steps. The neural network, parameterized by weights $\theta$, evaluates the current physical state to predict a continuous, cell-centered acceleration field $\mathbf{a}_{nn}^c = \mathcal{N}_\theta(S)$. Because the scalar variables and the horizontal velocities are staggered on the Arakawa C-grid, the cell-centered neural accelerations must be mapped to the cell interfaces to update the momentum components. We apply a simple arithmetic averaging to interpolate the continuous neural forcing to the respective velocity nodes
\begin{align*}
 &a_{u, i+1/2, j} = \frac{1}{2}\left(a_{u, i, j}^c + a_{u, i+1, j}^c\right),  \\ 
 &a_{v, i, j+1/2} = \frac{1}{2}\left(a_{v, i, j}^c + a_{v, i, j+1}^c\right).
\end{align*}
To maintain numerical stability and ensure the neural network does not arbitrarily force dry cells, the interpolated accelerations are modulated by the continuously differentiable wet--dry masking function. The mask is calculated at the cell interfaces by multiplying the wet fractions of the adjacent cell centers. The final, neurally-corrected velocity field $U^{m+1}$ is then updated via
\begin{align*}
&U^{m+1} = \tilde{U}^{m+1} + \Delta t \cdot a_u \cdot (M_{i,j} \cdot M_{i+1,j}),\\
&V^{m+1} = \tilde{V}^{m+1} + \Delta t \cdot a_v \cdot (M_{i,j} \cdot M_{i,j+1}).
\end{align*}
This fractional-step integration preserves the fundamental stability of the underlying momentum-conserved advection scheme while ensuring the exact gradients of the corrective forcing flow continuously back into the network weights.

\medskip
\noindent\textbf{Accelerated linear solvers.} The non-hydrostatic corrector requires solving the asymmetric system $Pq=Q$. Traditional approaches often rely on the Bi-CGSTAB algorithm paired with Incomplete Lower Upper (ILU) preconditioning. However, standard ILU factorization is sequential, and its lack of parallelism creates severe bottlenecks for accelerated linear algebra compilers. In \texttt{AegirJAX}, we adapt a solution strategy based on the domain dimensionality. For one-dimensional domains, the Poisson matrix reduces to a tridiagonal system, allowing us to bypass iterative solvers entirely in favor of an exact, differentiable, and parallelizable direct tridiagonal solver. For two-dimensional domains, we utilize a differentiable implementation of the Bi-CGSTAB algorithm. To avoid the sequential bottleneck of ILU, we employ a simpler, fully parallelizable Jacobi (diagonal) preconditioner: $M=\text{diag}(P)^{-1}$. Furthermore, matrix-vector products are evaluated using shifted array operations and slicing rather than sparse matrix indices, ensuring maximal computational throughput on GPUs.

\medskip
\noindent\textbf{Vectorized ghost cell padding.} Throughout the explicit predictor step and the momentum-conserved advection scheme, traditional models loop over interior domain indices and apply specialized conditional logic at the boundaries. In \texttt{AegirJAX}, all spatial gradients, flux limiters, and upwind evaluations are formulated using whole-array slicing and contiguous array padding operations. Boundary conditions, such as closed reflective walls or continuous open boundaries, are enforced purely by specifying the mathematical mode of the boundary padding. This approach eliminates the need for non-differentiable control flow routines, ensuring that the computational graph remains static and structured for JAX' just-in-time compilation.

\section{Applications to wave propagation problems}\label{sec:NumericalExperiments}

In this section we present the capabilities of the differentiable solver framework developed in Section~\ref{sec:Theory}, by applying it to a wide variety of classical and novel benchmarks of relevance for wave propagation and inundation problems. Technical details for the specific domain definitions and time-stepping, training procedures and neural network architectures used for each example can be found in Appendix~\ref{sec:Appendix}.

\subsection{Model discovery and physical correction}

While numerical solvers derived from first principles are foundational to coastal engineering, they inherently suffer from truncation errors and the omission of complex, sub-grid physical processes. Formulations are often simplified, such as assuming hydrostatic pressure or depth-integrating the vertical dimension to maintain computational tractability. When applied to non-stationary regimes, these approximations result in phase shifts, amplitude damping, and missing harmonic generation~\cite{mads91a}. Differentiable programming offers a pathway to discover and re-inject these missing physics natively within the solver, via the computation of the neural network based acceration field $\mathbf{a}_{\rm nn}$. By back-propagating the error between simulated states and experimental gauge observations through the unrolled temporal simulation, a neural network can be trained to act as a localized forcing term, dynamically predicting the momentum corrections required to bridge the gap between the simplified base solver and the true physical reality.

Before detailing the specific benchmarks, it is critical to distinguish the data-driven objectives of these experiments based on data availability. Because we rely on historical, large-scale laboratory and real-world datasets, the volume of available experimental runs is inherently sparse. For benchmarks restricted to single events or isolated wave conditions, the neural network cannot learn a universally applicable sub-grid parameterization. Instead, it effectively acts as a specialized, regime-specific curve-fitting algorithm. This limitation is not a shortcoming of the \texttt{AegirJAX} framework, but rather a direct consequence of data scarcity. However, as we will demonstrate in the final benchmark, when sufficient variations of wave conditions are available, the differentiable solver transcends trajectory memorization to learn generalized, interpolatable physical corrections.

\subsubsection{Dispersive correction in the Beji--Battjes benchmark problem}\label{sec:BejiBattjes}

The Beji and Battjes (1993) benchmark~\cite{beji93a} is a classic test case for dispersive wave propagation. As a sinusoidal wave train propagates over a submerged trapezoidal bar, the shoaling process induces strong nonlinearities. As the wave passes into the deeper water behind the bar, it decomposes into higher-frequency dispersive tails (super-harmonics). Standard one-layer depth-averaged non-hydrostatic models struggle to accurately resolve these short free surface waves without the addition of multiple vertical layers to capture the complex vertical velocity profiles~\cite{zijl08a}. 

To correct this limitation of the base solver, we couple \texttt{AegirJAX} with a one-dimensional dilated convolutional residual neural network designed to predict a conservative pseudo-pressure potential $\phi(x)$. The acceleration correction is then derived via the spatial gradient, $a_{\rm nn} = \partial\phi/\partial x$, mathematically preventing the artificial injection of net momentum. To ensure the network learns the underlying dispersive physics rather than arbitrary noise, it is fed a physically motivated feature space. Alongside standard kinematic variables, the network receives the local Froude number and higher-order spatial derivatives of the free surface, most notably the Laplacian $\partial^2 \zeta/\partial x^2$ and the third spatial derivative $\partial^3 \zeta / \partial x^3$. These terms act as dispersion triggers, providing the network with the mathematical foundation equivalent to Boussinesq-type dispersive equations. Furthermore, to prevent the network from generating non-physical drift in quiescent water prior to the wave's arrival, an input gating mechanism is applied to the features, activating the neural network only in the presence of local wave kinetic energy or surface elevation deviations.

\begin{figure}[!ht]
    \centering
    \includegraphics[width=\linewidth]{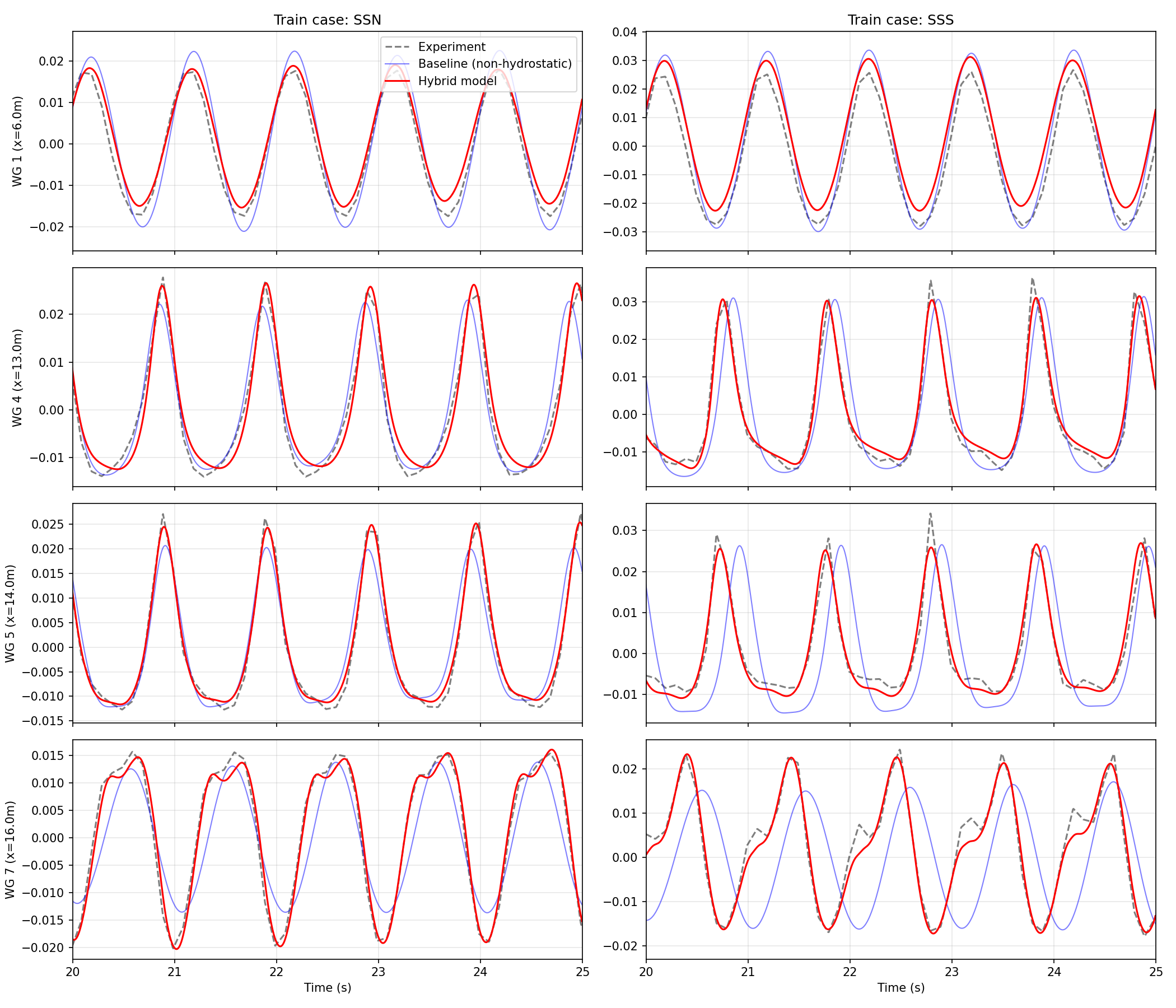}
    \caption{Correction of missing dispersive physics in the Beji--Battjes benchmark~\cite{beji93a} The baseline hydrostatic model (blue) fails to capture the short-wavelength super-harmonics behind the submerged bar. By utilizing higher-order spatial derivatives as dispersion triggers, the neural-augmented solver (red) learns a regime-specific forcing that accurately recovers the experimental wave amplitudes and phases (black) for both non-breaking and spilling wave cases.}
    \label{fig:BejiBattjesCorrectionResults}
\end{figure}

Because this benchmark relies on isolated experimental runs, the training manifold lacks the dense, diverse variations of wave conditions required for a neural network to learn a universally applicable sub-grid parameterization. The objective here is strictly regime-specific physical correction. As illustrated in Figure~\ref{fig:BejiBattjesCorrectionResults}, the neural-augmented solver successfully learns to inject the missing physics for these exact scenarios, correcting the waveforms to better match the experimental gauge data, leading to an overall mean squared error reduction of 85.9\% (from $4.96\cdot 10^{-6}$ to $7.01\cdot 10^{-7}$) for the Sinusoidal Short Non-breaking wave (SSN) case and a 90.8\% reduction (from $2.46\cdot 10^{-5}$ to $2.26\cdot 10^{-6}$) for the Sinusoidal Short Spilling wave (SSS) case, without the need to resort to a more complicated multi-layer non-hydrostatic solver. In particular, the model perfectly captures the complex non-hydrostatic wave dispersion and the precise phase of the super-harmonics generated immediately after the wave passes over the submerged mountain.

\subsubsection{Correction of missing physics in the Monai Valley benchmark}

To further test the robustness of the differentiable solver paradigm, we apply the two-dimensional scalar potential correction to the Monai Valley benchmark. Based on the 1993 Hokkaido Nansei--Oki earthquake, this benchmark represents a complex real-world bathymetry characterized by a narrow coastal valley where the incident tsunami resulted in a maximum run-up of 31.7 meters. Modeling the Monai Valley tsunami requires capturing extreme wave shoaling, reflection, and inundation over a highly irregular grid. 

To preserve the fundamental stability of the shallow-water equations, we enforce an irrotational neural network forcing. As in the previous case, a dilated convolutional residual neural network is designed to predict a single-channel scalar potential $\phi(x,y)$ from which the acceleration field is derived via the spatial gradient, $\mathbf{a}_{\rm nn} = \nabla \phi$. The input feature space provided to the network includes localized depths, velocities, surface gradients, bed slopes, and flow vorticity. Crucially, to model the missing Boussinesq-type dispersive effects, the feature space is augmented with the surface elevation Laplacian $\nabla^2 \zeta$, providing the local curvature information strictly necessary for the network to learn accurate phase-shifting corrections. To ensure stability at the moving waterline, the neural forcing is modulated by a continuously differentiable depth mask. Furthermore, to eliminate non-physical drift in still water while allowing the network to force complex wave collisions, an input gate with a threshold deadzone is applied to the input features. For this experiment, the baseline model is a purely hydrostatic solver on a coarsened grid, using a down-sampling stride of 4. The extreme gradients of the real-world bathymetry inherently lower the numerical stability of the baseline explicit scheme. Injecting neural forcing into such an environment risks exciting high-frequency instabilities. To counteract this, we introduce a Total Variation (TV) regularization penalty into the loss function during training,
\[
\mathcal{L}_{\rm TV} = \lambda_{\rm TV} \sum \left( (\nabla_x \mathbf{a}_{nn})^2 + (\nabla_y \mathbf{a}_{nn})^2 \right),
\]
where $\lambda_{\rm TV}$ heavily penalizes spatial noise in the predicted forcing fields.

\begin{figure}[!ht]
    \centering
    \includegraphics[width=\linewidth]{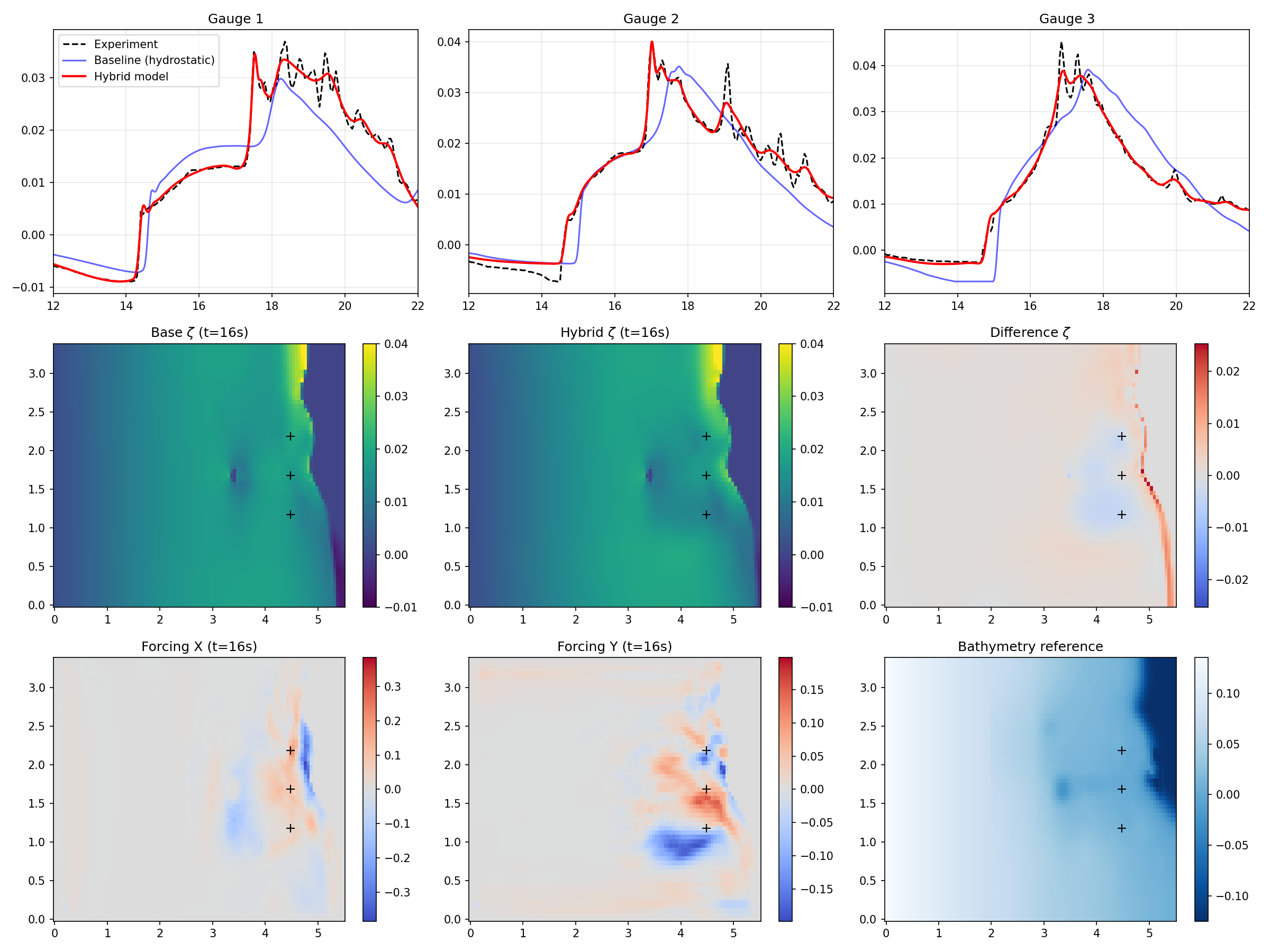}
    \caption{ Neural-augmented correction of the Monai Valley benchmark. \textit{Top row:} Wave gauge time-series comparing the baseline hydrostatic solver, the hybrid model, and experimental ground truth. \textit{Middle and bottom rows:} Snapshots of the surface elevation differences and the corresponding spatial neural momentum corrections $\mathbf{a}_{\rm nn}$ at $t=16$ s. The network injects localized forcing to correct the wave routing and secondary reflections missed by the coarse-grid baseline.}
    \label{fig:MonaiValleyResults}
\end{figure}

Because there is only a single historical/laboratory tsunami trajectory available for this bathymetry, the neural network acts again as an advanced, physics-constrained curve-fitting mechanism. Figure~\ref{fig:MonaiValleyResults} highlights the performance of the differentiable hybrid model. The baseline coarse-grid hydrostatic solver fails to capture the correct amplitude of the primary wave peak and completely misses the complex secondary reflections triggered by the valley's geometry. By integrating the neural scalar potential, the hybrid model successfully corrects the wave profiles across all three primary gauges, yielding an improvement of 94.2\% in the mean squared error, from $2.95\cdot 10^{-5}$ in the hydrostatic base model to $1.71\cdot 10^{-6}$ in the hybrid model that learned the neural network corrected velocity $\mathbf{a}_{nn}$. The visualization of the forcing field demonstrates that the network dynamically activates as the wave interacts with the coastline, generating smooth, TV-regularized accelerations that correct the coarse-grid routing without destabilizing the underlying shallow-water integration.

\subsubsection{Correction of missing physics in the Conical iIsland benchmark}\label{sec:ConicalIsland}

While the previous benchmarks successfully discovered regime-specific forcing, achieving true machine learning generalizability requires learning physical corrections that apply to unseen data. The Conical Island benchmark, based on the large-scale laboratory experiments of Briggs et al. (1995)~\cite{brig95a}, provides the necessary dataset to test this, since there are a total of three runs that were carried out with varying incident waves. In this experiment, solitary waves interact with a truncated Conical Island, generating complex wave--wave interactions and rotational wake structures. As for the Monai Valley benchmark correction in the previous subsection, we task the differentiable framework with correcting a cheap, purely hydrostatic base solver so that it accurately reproduces the complex three-dimensional flow physics observed in the physical experiment.

Crucially, to evaluate the generalization capabilities of the hybrid solver, we train the neural network jointly on the edge cases of the dataset: Case A (small wave, $A/h \approx 0.045$) and Case C (large wave, $A/h \approx 0.181$). The loss function minimizes the spatio-temporal mean squared error at four gauge locations around the island over the active wave-interaction window. The trained hybrid model is then evaluated zero-shot on the intermediate Case B ($A/h \approx 0.096$). 

\begin{figure}[!ht]
    \centering
    \includegraphics[width=\linewidth]{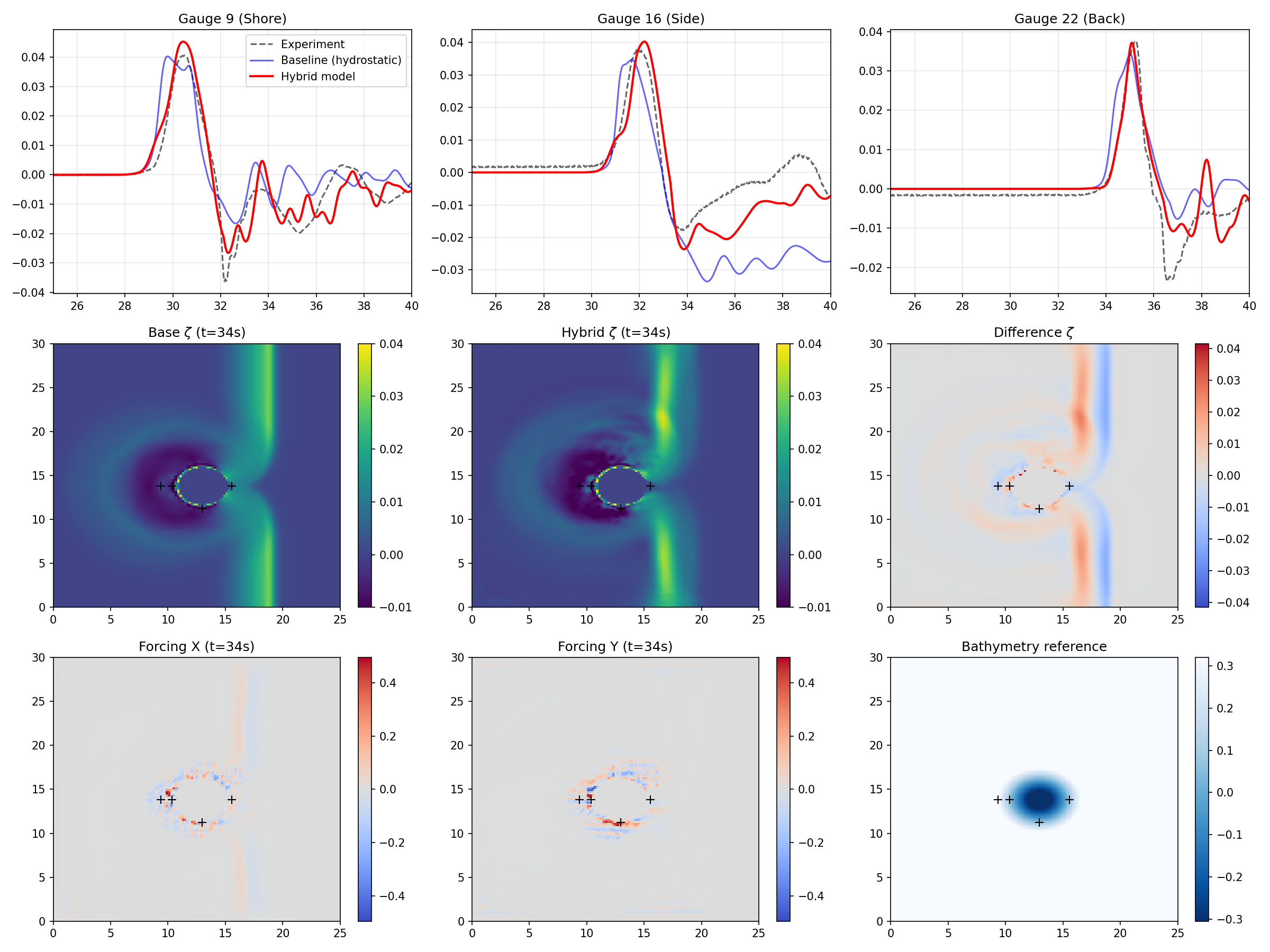}
    \caption{Zero-shot generalization of the neural correction on the Conical Island benchmark. The hybrid model, trained only on extreme wave amplitudes, is evaluated on an unseen intermediate wave, $A/h \approx 0.096$. The neural-augmented solver successfully interpolates the missing physics, significantly reducing phase errors and amplitude under-predictions compared to the coarse-grid hydrostatic baseline.}
    \label{fig:ConicalIslandResults}
\end{figure}

As shown in Figure~\ref{fig:ConicalIslandResults}, the baseline hydrostatic model suffers from significant phase errors and under-predicts the wave amplitudes. The physics-conditioned neural correction successfully interpolates the missing physics for the unseen wave, aligning the phase and amplitude with the experimental data and yielding an 74.4\% improvement in the mean squared error from the baseline MSE of $3.67\cdot 10^{-5}$ to $9.41\cdot 10^{-6}$. This confirms that when provided with even a minimal envelope of diverse training data, \texttt{AegirJAX} does not simply memorize trajectories, but rather learns a generalized sub-grid parameterization of the underlying fluid dynamics.

We should like to stress here that this benchmark still constitutes a rather pathological machine learning task: Having \textit{two} data samples available to train a neural network based correction of a \textit{single} unseen test example. A wider variety of training datasets, would almost surely further improve the generalization capabilities of end-to-end differentiable numerical solvers such as \texttt{AegirJAX}.

\subsection{Inverse design and topological optimization}

Traditional coastal engineering design relies heavily on trial-and-error, heuristic guidelines, or computationally expensive gradient-free optimization methods such as genetic algorithms~\cite{babo00a}. By utilizing \texttt{AegirJAX}, we can reframe structural design as a formal inverse problem. Because the fluid solver is end-to-end differentiable, the bathymetry and boundary conditions can be treated as trainable parameters. We demonstrate this capability through two distinct paradigms: the placement of a parameterized rigid body, and the continuous topological distribution of a fixed volume of breakwater material. Since no real world datasets are available for these benchmarks, we will be working with synthetic data in the sequel.

\subsubsection{Optimization of a fixed gate position}

\begin{figure}[!ht]
    \centering
    \includegraphics[width=\linewidth]{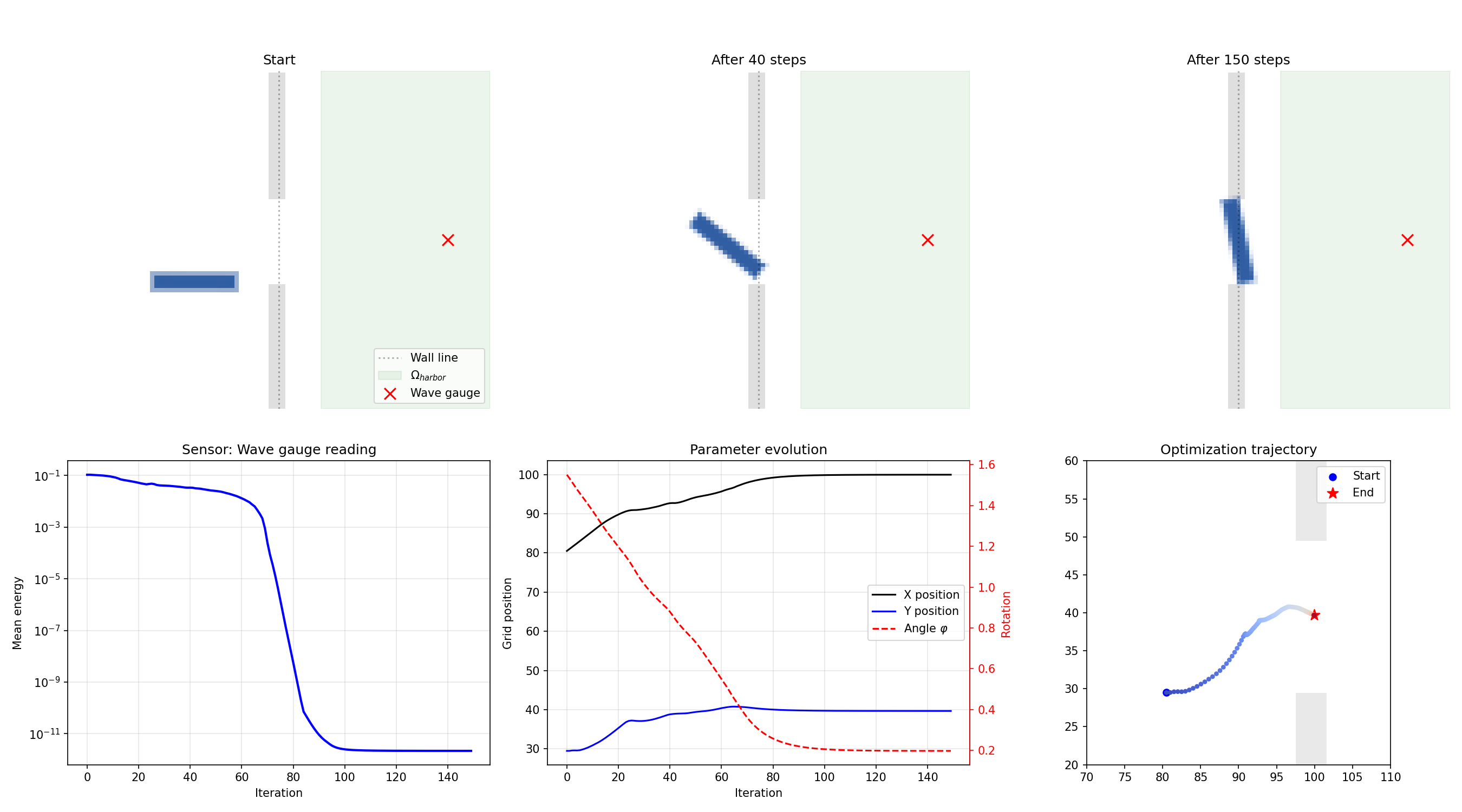}
    \caption{Optimization of a rigid breakwater gate to minimize harbor inundation. \textit{Top:} Initial misaligned state, intermediate optimization step, and final converged gate position. \textit{Bottom:} Convergence history of the wave gauge energy, the time series of the optimized variables and the top-down spatial trajectory of the gate's center coordinates guided by analytical gradients into the static wall gap.}
    \label{fig:GatePlacementBenchmark}
\end{figure}

In the first experiment, we consider the task of positioning a rigid, rectangular breakwater gate to protect a harbor from an incident solitary wave. The domain consists of a static sea wall with a defined gap. The objective is to determine the optimal center coordinates $(x_c, y_c)$ and rotation angle $\varphi$ of the gate to minimize the total kinetic energy entering the protected harbor zone. The primary challenge in rigid-body optimization on a Eulerian grid is that moving a discrete object creates step-changes in the bathymetry, yielding zero gradients almost everywhere and infinite gradients at the edges. To make the rigid body differentiable, we employ a continuous rasterization technique based on a signed distance field. The distance $d(x,y)$ from any grid cell to the boundary of the rotated rectangle is computed, and the presence of the structure is mapped onto the grid using a steep sigmoid function,
\[
M(x,y) = \sigma\big(-k \cdot d(x,y, x_c, y_c, \varphi)\big),
\]
where $k$ controls the sharpness of the structural boundary. The physical water depth is then modified as $h_{\rm sim} = h_{\rm initial} - M(x,y)H_{\rm gate}$. Because $M(x,y)$ provides a continuous, non-zero gradient gradient field both inside and outside the structure, the optimizer can determine which direction to move and rotate the gate to block the incoming wave energy. The loss function is defined as the spatio-temporal sum of kinetic energy in the harbor,
\[
\mathcal{L} = \sum_{t} \sum_{(x,y) \in \Omega_{\rm harbor}} \frac{1}{2} h \big(U^2 + V^2\big).
\]
As shown in Figure~\ref{fig:GatePlacementBenchmark}, the optimization algorithm is able to translate and rotate the initial misaligned gate seamlessly across the domain. The gradients back-propagated through the hydrodynamic solver successfully guide the gate to park exactly within the gap, confirming that \texttt{AegirJAX} can optimize discrete structural parameters via continuous spatial relaxations.

\subsubsection{Topological optimization for optimal breakwater design}

While rigid body positioning optimizes a predefined shape, topology optimization seeks to discover the optimal structural shape itself from scratch. In this experiment, the goal is to protect a specific nearshore facility located on a sloping beach from an incident solitary wave, using a fixed volume of breakwater material. We formulate this problem using an approach inspired by the Solid Isotropic Material with Penalization (SIMP) method commonly used in structural mechanics~\cite{bend13a}. Rather than optimizing depth directly, we define a latent field of unconstrained logits $\alpha(x,y)$ across the allowable building zone. These logits are transformed into a normalized density field $\rho \in [0, 1]$ using a sigmoid activation, followed by a two-dimensional Gaussian smoothing convolution. The smoothing step is critical as it enforces a spatial correlation that prevents the optimizer from generating non-physical, checkerboard-like scattered pillars, instead encouraging contiguous, buildable breakwater structures. 

To penalize the optimizer for creating porous, shallow mounds that do not effectively block waves, the physical added height $\Delta h$ is scaled cubically relative to the density,
\[
\Delta h(x,y) = \rho(x,y)^3 H_{\rm max}.
\]
This cubic penalty forces the optimization process to converge toward binary states, that is, either no material ($\rho \approx 0$) or full-height walls ($\rho \approx 1$). The modified bathymetry is then passed to the solver. The optimization objective balances the potential energy of the wave in the facility zone against a fixed material budget,
\[
\mathcal{L} = \sum_{t} \sum_{(x,y) \in \Omega_{\rm facility}} \zeta^2 + \lambda \left| \frac{V_{\rm current} - V_{\rm target}}{V_{\rm target}} \right|,
\]
where $V_{\rm current} = \sum \rho H_{\rm max} \Delta x \Delta y$ is the total volume of material used, $V_{\rm target}$ is the allowable volume budget, and $\lambda$ is a penalty weighting factor.

\begin{figure}[!ht]
    \centering
    \includegraphics[width=\linewidth]{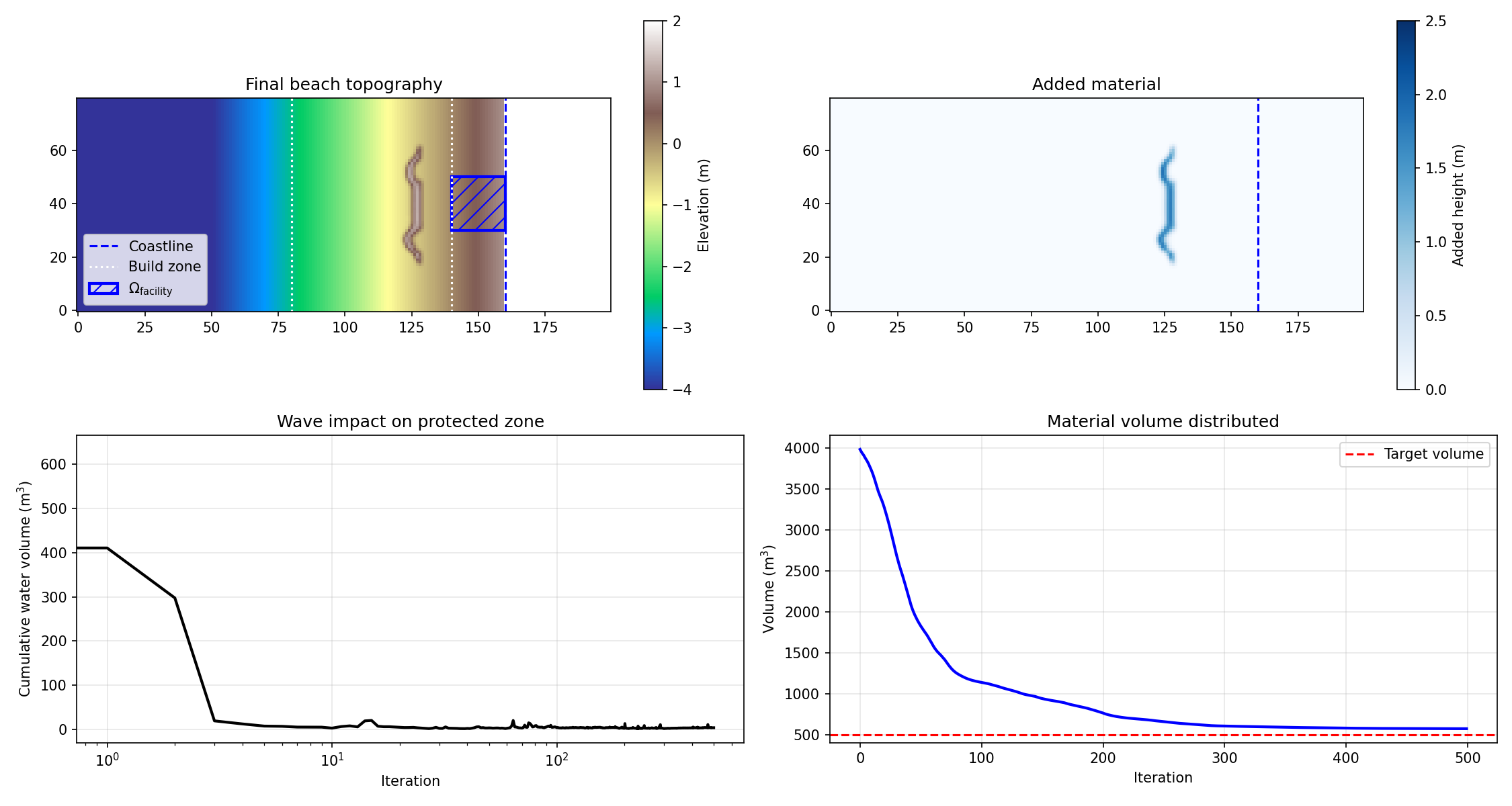}
    \caption{ Continuous topology optimization for breakwater design under a strict material volume constraint. \textit{Top:} The initial sloping beach topography and the final optimally distributed breakwater material $\Delta h$. \textit{Bottom:} The optimization history demonstrating the minimization of wave impact energy on the facility \textit{(left)} while adhering to the target material volume budget \textit{(right)}. The optimizer autonomously discovers that a curved offshore berm optimally refracts the incident wave.}
    \label{fig:BreakwaterDesign}
\end{figure}

As illustrated in Figure~\ref{fig:BreakwaterDesign}, through gradient descent, the latent logit field evolves from a uniform distribution of noise into a specialized breakwater geometry. The optimizer autonomously discovers that distributing the limited material into a continuous, curved offshore berm optimally refracts and reflects the incident wave energy away from the target facility, satisfying the volume constraint while minimizing inundation.

\subsection{Active control of hydrodynamic systems}

Active control of coastal structures, such as dynamic breakwaters, wave-makers, or pneumatic wave energy converters, presents a substantial engineering challenge~\cite{ring14a}. The control policy must act in real-time based on limited sensor data and account for the complex, nonlinear temporal dynamics of wave propagation. Traditional approaches either rely on simplified linear control theory or utilize model-free reinforcement learning~\cite{sutt98a}. In recent years, deep reinforcement learning has been increasingly explored for the active control of wave energy converters and coastal structures~\cite{ande20a,faed17a}. Model-free reinforcement learning, however, is sample-inefficient because it has to learn the underlying fluid physics entirely through trial and error~\cite{holl20a}. By utilizing \texttt{AegirJAX}, we can employ a \textit{policy-in-the-loop} training paradigm. Because the entire unrolled fluid simulation is a differentiable computational graph, we can train a neural network policy using exact analytical gradients computed via back-propagation through time, directly from the downstream hydrodynamic loss to the network weights.

To demonstrate this, we develop a stateful recurrent neural network policy for active wave cancellation. The computational domain consists of three distinct zones: a sensor array, an active wave-maker (actuator) with a Gaussian spatial influence profile, and a protected harbor zone. The objective of the neural policy is to read the incoming wave states from the sensor array and trigger the actuator to generate a destructive interference wave, thereby minimizing the wave energy entering the harbor. Because wave propagation is inherently temporal, a static feed-forward network mapping current sensor readings to current actions is insufficient, and the policy must anticipate the wave's arrival based on its propagation history. Therefore, we parameterize the control policy using a recurrent neural network based on a Gated Recurrent Unit (GRU) architecture~\cite{cho14a}. At each simulation time step, the recurrent neural network ingests the local free-surface elevation $\zeta$ and horizontal velocity $u$ from the sensor array. It updates its internal hidden memory state and outputs a bounded scalar force command to the physical actuator.

To ensure the policy learns a generalized wave-cancellation strategy rather than simply memorizing a single trajectory, the recurrent neural network is trained on a batch of incident solitary waves with varying initial amplitudes and starting positions. The loss function is defined as the sum of the squared free-surface elevation (a proxy for potential energy) entering the protected harbor, augmented by an $L_2$-regularization penalty on the actuator to discourage unnecessarily violent control actions,
\[
\mathcal{L} = \sum_{t} \sum_{x \in \Omega_{\rm harbor}} \zeta(x,t)^2 + \lambda_{\rm effort} \sum_{t} F_{\rm actuator}(t)^2.
\]
The gradients of this combined loss are back-propagated through the non-hydrostatic Boussinesq solver and into the recurrent neural network's weights. The optimizer then successfully discovers a control strategy that waits for the wave to reach the optimal position before activating the actuator in opposition to the wave's momentum.

\begin{figure}[!ht]
    \centering
    \includegraphics[width=\linewidth]{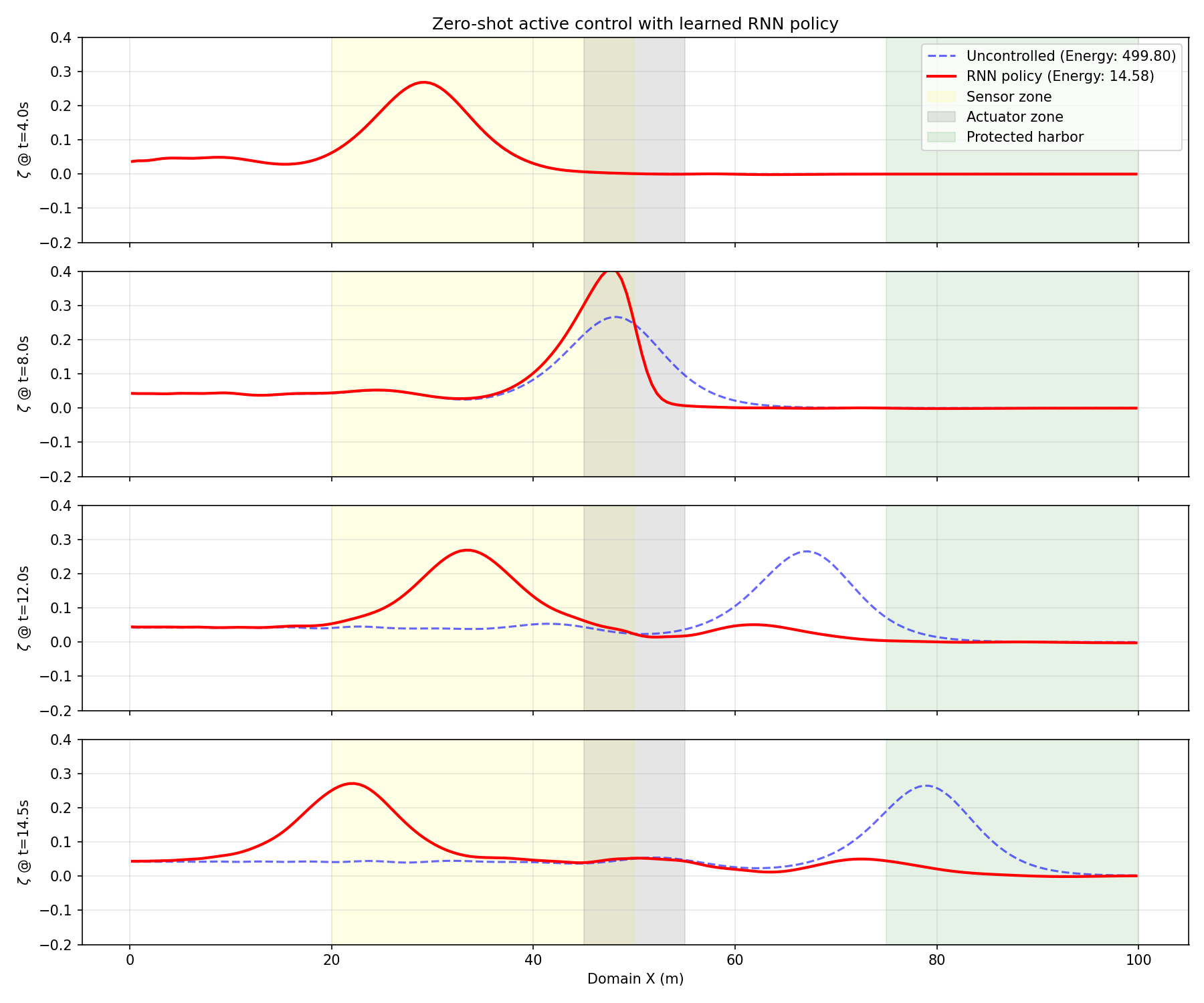}
    \caption{Zero-shot active wave cancellation using a policy-in-the-loop recurrent neural network. The learned policy reads incoming wave states from an upstream sensor array (yellow region) and commands a parameterized actuator (gray region). The RNN accurately anticipates the unseen wave's phase and amplitude, applying a destructive interference force that reduces the total energy entering the protected harbor (green region) by over 97\%.}
    \label{fig:ActiveControlResults}
\end{figure}

To validate the robustness of the learned policy, it is evaluated zero-shot on an unseen incident wave with amplitude $A=0.28$ m outside the training distribution. Figure~\ref{fig:ActiveControlResults} illustrates the time evolution of this zero-shot test. In the uncontrolled baseline simulation, the solitary wave passes freely through the domain, delivering a total cumulative energy metric of 499.80 to the harbor. When the trained recurrent neural network policy is engaged, it accurately anticipates the wave's phase and amplitude, actuating the fluid to drastically dampen the transmitted wave. The active control policy reduces the total harbor energy to 14.58, an energy reduction of over 97\%, demonstrating the high efficacy of directly coupling recurrent neural networks with differentiable hydrodynamical solvers.

\subsection{Source inversion and parameter estimations}

A fundamental challenge in coastal and off-shore engineering is inferring unobservable physical fields, such as complex underwater topography or the dynamics of submarine mass failures, from limited surface observations. Traditional discrete adjoint methods often struggle with the ill-posed nature of these problems, particularly when the forward model involves highly dispersive or nonlinear moving-boundary physics~\cite{gile03a, monn16a}. While continuous and recursive adjoint models have been developed for higher-order spectral wave simulations to reconstruct bathymetry, deriving these terms requires expanding the water wave equations to specific perturbation orders, rendering them computationally inflexible when boundary conditions change~\cite{wu23a}. By leveraging the differentiable programming paradigm, \texttt{AegirJAX} allows us to formulate these inversions as physics-constrained machine learning tasks, where continuous neural representations regularize the parameter space and exact gradients flow directly from downstream wave gauges back to the source mechanisms.

\subsubsection{Estimation of an unknown fixed bottom topography}

In this experiment, we tackle the classic bathymetry inversion problem: recovering the spatial profile of the submerged trapezoidal bar from the Beji--Battjes benchmark discussed in Section~\ref{sec:BejiBattjes} from the free-surface elevation time series recorded by downstream wave gauges.

Directly optimizing the depth value at every computational grid point is an under-determined problem that typically results in non-physical, high-frequency spatial noise. To enforce a spatial inductive bias, we parameterize the continuous bathymetry field using a multi-layer perceptron. The network maps the spatial coordinate $x$ to the water depth: $h(x) = \mathcal{N}_\theta(x)$. The optimization task is then reduced to finding the optimal neural network weights $\theta$. The loss function is defined as the mean squared error between the simulated and observed surface elevations at the specific gauge locations, augmented by an $L_2$ spatial smoothness penalty and a volume constraint to anchor the deep-water floor,
\[
\mathcal{L} = \sum_{g=1}^{N_{\rm gauges}} \sum_{t} \big(\zeta_{\rm sim}(x_g, t) - \zeta_{\rm obs}(x_g, t)\big)^2 + \lambda_{\rm smooth} \sum_{x} \left( \frac{\partial h(x)}{\partial x} \right)^2 + \lambda_{\rm vol} \sum_{x} (h_0 - h(x))^2.
\]
Starting from an initial flat-bottom guess, the optimizer back-propagates the hydrodynamic loss through the non-hydrostatic solver. 

\begin{figure}[!ht]
    \centering
    \includegraphics[width=\linewidth]{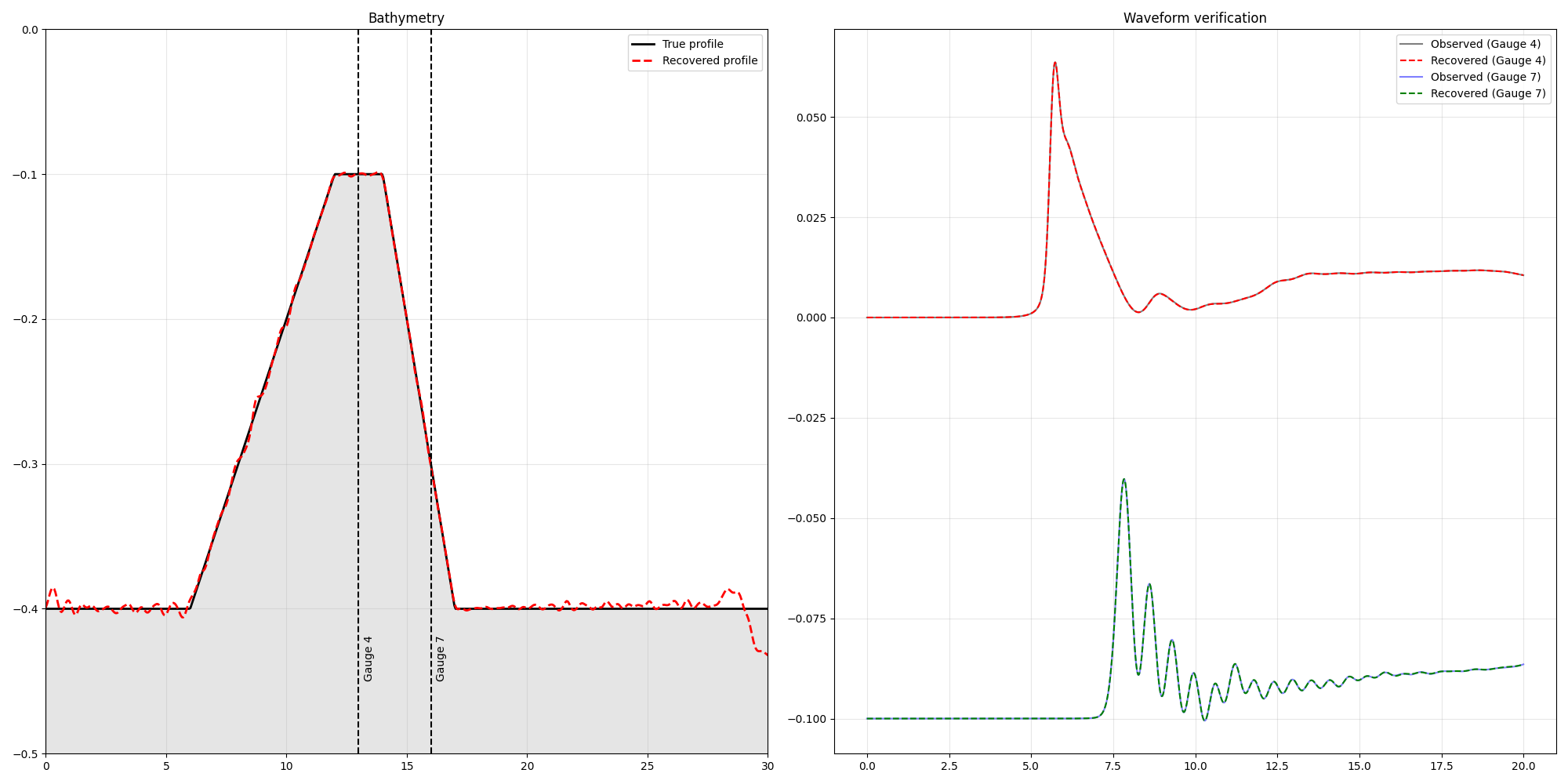}
    \caption{Recovery of the submerged trapezoidal bathymetry~\cite{beji93a} from downstream wave gauge observations. \textit{Left:} A continuous neural network parameterization accurately recovers the front and lee slopes of the bar. The deviation on the far right reflects the physical ill-posedness of the unmonitored shadow zone. \textit{Right:} The resulting wave time-series demonstrating precise agreement with the downstream observations.}
\label{fig:BejiBattjesTopographyRecovered}
\end{figure}

Figure~\ref{fig:BejiBattjesTopographyRecovered} displays the convergence of the loss and the final recovered bathymetry profile. The network successfully recovers the complex geometry of the submerged bar, including the precise slopes of the front and lee faces. Notably, there is a distinct discrepancy between the true profile and the neural network-recovered profile on the far right side of the domain. This divergence is a feature of the physical formulation rather than a numerical artifact. Because there are no wave gauges situated past this point to record the wave transformation, the wave state traversing that region provides no upstream gradient signal to the optimizer. The mathematical ill-posedness of the region is reflected in the optimization process in that the solver correctly identifies that the topography in the unmonitored shadow zone is unconstrained by the downstream observations.

\subsubsection{Estimation of an unknown submarine landslide}

While recovering static bathymetry demonstrates spatial inversion, assessing tsunami hazards often requires inverting for dynamic, moving sources. In this experiment, we recover both the kinematics and the spatial deformation shape of a submerged submarine landslide on a static sloping bottom topography entirely from the downstream waves it generates. Following the benchmark established in~\cite{gril05a}, the landslide is modeled as a solid mass accelerating down a slope. The temporal motion of the slide's center of mass, $s(t)$, is governed by the exact log-cosh kinematic law
\[
s(t) = u_t t_0 \ln\left[\cosh\left(\frac{t}{t_0}\right)\right],
\]
where $u_t$ is the terminal velocity and $t_0$ is the characteristic time of acceleration. The time-dependent bathymetry that forces the fluid solver is defined as $h(x,t) = h_{\rm static}(x) - S(x - s(t))$, where $S(\xi)$ is the shape function of the slide in its moving reference frame. Instead of assuming a standard Gaussian or rigid elliptical shape, we treat the spatial distribution of the slide as an unknown. We parameterize $S(\xi)$ using a multi-layer preceptron neural network with Fourier feature embeddings, while simultaneously treating the kinematic width parameter $b$, parameterized as $\exp(\log b)$, and the base thickness parameter $T$, parameterized as $\exp(\log T)$, to enforce positivity, as trainable scalars. The optimization process jointly updates the weights of the shape network and the scalar kinematic parameters by back-propagating the error from the downstream wave gauges through the fully unrolled temporal sequence of the moving-boundary fluid solver. 
\begin{figure}[!ht]
    \centering
    \includegraphics[width=\linewidth]{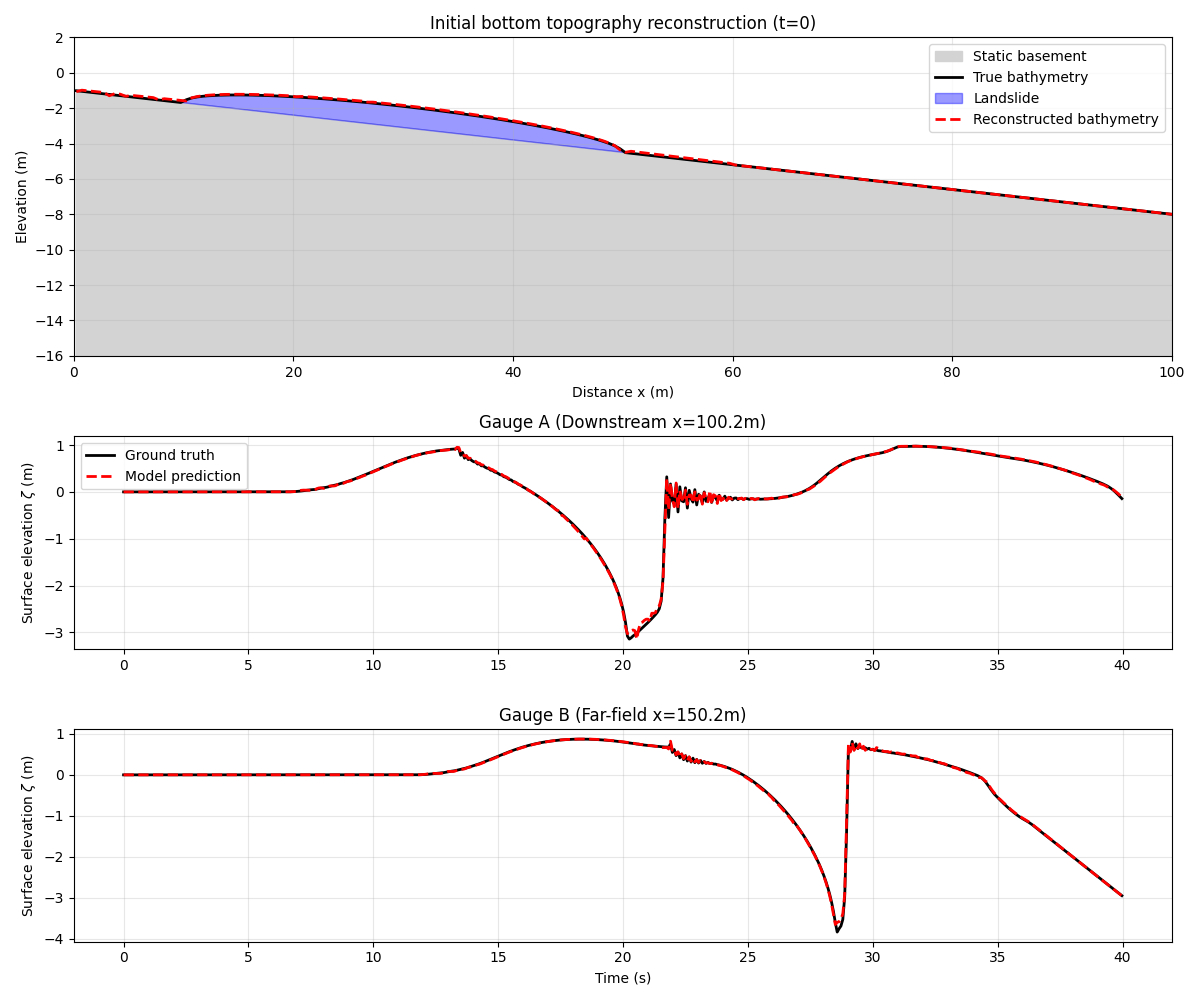}
    \caption{Joint inversion of submarine landslide kinematics and spatial deformation. \textit{Top:} Comparison of the true and reconstructed initial bottom topography at $t=0$. \textit{Middle and bottom:} Reconstructed surface elevation time-series at near-field and far-field gauges. By back-propagating the downstream error, \texttt{AegirJAX} successfully recovers both the spatial footprint of the mass and its log-cosh acceleration profile.}
    \label{fig:LandslideRecoveryResults}
\end{figure}

As illustrated in Figure~\ref{fig:LandslideRecoveryResults}, the joint optimization successfully matches the complex, dispersive wave trains recorded at the sensors. By recovering both the spatial footprint of the deformation and the temporal scale of its acceleration, this experiment highlights the capability of \texttt{AegirJAX} to seamlessly fuse parameter estimation and functional regression in moving-boundary hydrodynamic environments.

\subsection{Coupling multi-physics for passive tracer source localization}

To further demonstrate the extensibility of the differentiable \texttt{AegirJAX} framework, we consider the environmental engineering challenge of localizing a passive tracer source, such as a chemical spill or pollutant leak, from sparse downstream sensor observations. Traditionally, framing this as a gradient-based inverse problem using the adjoint-state method is problematic. Deriving the continuous adjoint for the advection-diffusion equation requires running diffusion backwards in time, a mathematically ill-posed operation that typically amplifies high-frequency spatial noise. By leveraging \texttt{AegirJAX}'s differentiable programming paradigm, we bypass this limitation by computing the exact gradients of the discrete forward operations instead.

We extend the core hydrodynamic solver by coupling it to the depth-integrated advection-diffusion equation,
\[
\frac{\partial C}{\partial t} + \mathbf{U} \cdot \nabla C = K \nabla^2 C + S(x,y,t),
\]
where $C$ is the tracer concentration, $K$ is the diffusion coefficient, and $S(x,y,t)$ is the injection source term. This new physical kernel can be implemented in fewer than twenty lines of code using an upwind scheme for advection and central differences for diffusion, perfectly integrating with the existing predictor--corrector time-marching loop without breaking the computational graph.

For the experimental setup, we revisit the Monai Valley benchmark setup. An incident tsunami wave inundates the coastal valley, generating a dynamic, non-linear velocity field. A synthetic ground-truth chemical spill is initiated directly upstream of the small island. As the tsunami wave washes over the domain, the tracer plume is advected downstream, bifurcating around the island to create a heterogeneous, complex concentration signal at the three downstream wave gauges.

The optimization task is to recover the exact continuous spatial coordinates $(x_c, y_c)$ and the total injection mass $M$ of the spill, given only the concentration time-series recorded by the three gauges. To ensure the gradients flow smoothly from the Eulerian grid into the continuous coordinate parameters, the point source is rasterized via a differentiable spatial Gaussian. The mass parameter is constrained to be strictly positive using a softplus activation. The loss function is defined as the mean squared error between the simulated and observed sensor concentrations.

\begin{figure}[!ht]
    \centering
    \includegraphics[width=\linewidth]{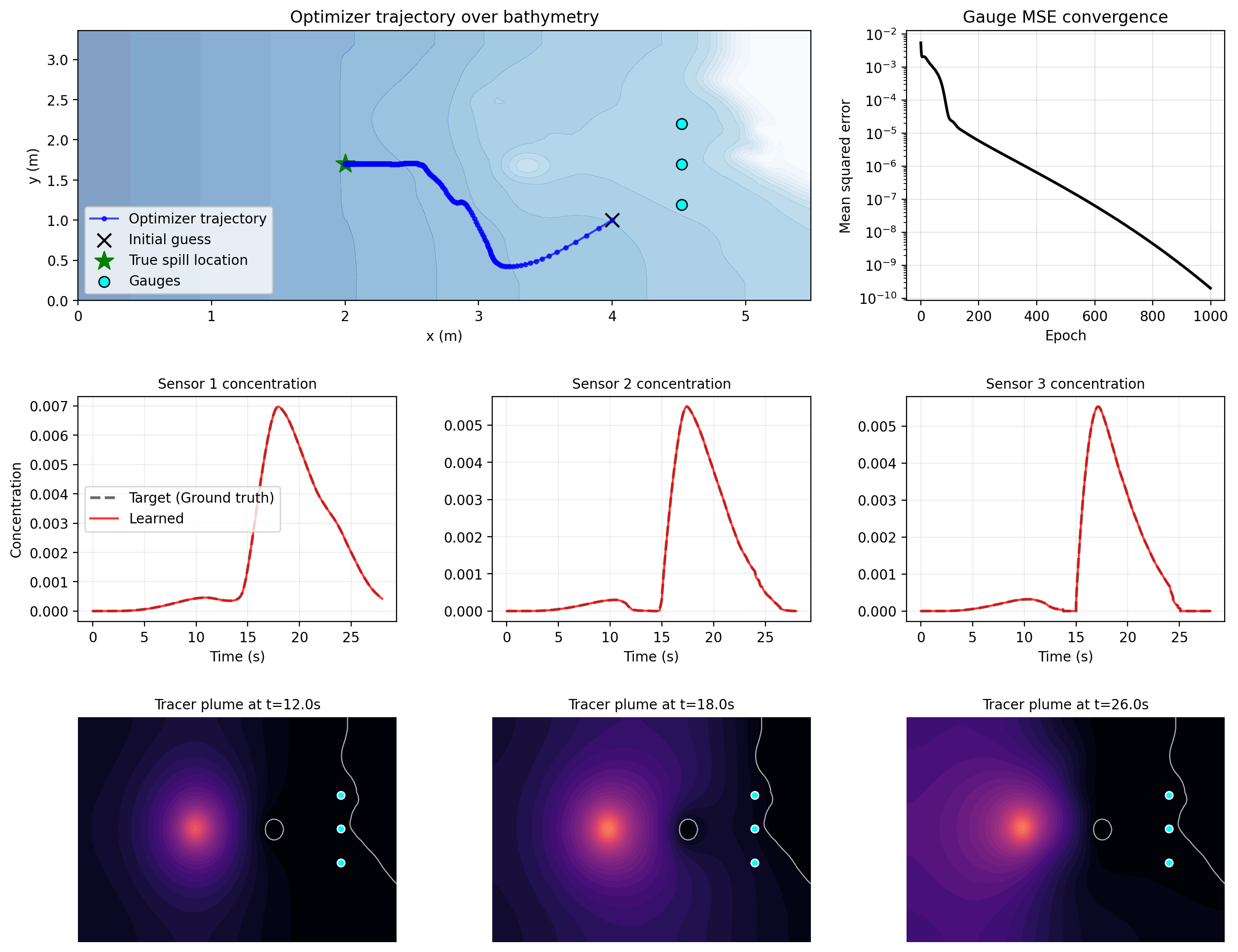}
    \caption{Inverse source localization of a passively advected tracer in the Monai Valley domain. \textit{Top:} The spatial trajectory of the source optimizer navigating the bathymetry to the ground-truth location, alongside the convergence of the gauge mean squared error. \textit{Middle:} The recovered concentration time-series at three downstream sensors. \textit{Bottom:} Snapshots of the simulated tracer plume advecting through the complex velocity field.}
    \label{fig:TracerInversion}
\end{figure}

As illustrated in Figure~\ref{fig:TracerInversion}, the optimizer is initialized with a severely misaligned guess on the opposite side of the bay. Driven entirely by the analytical gradients back-propagated through both the tracer transport and the underlying shallow-water hydrodynamics, the optimizer successfully navigates the complex bathymetry. The trajectory plot demonstrates that the optimizer correctly interprets the upstream gradients generated by the wave diffraction around the island, successfully guiding the source coordinates to the exact ground-truth location and recovering the correct injection mass.

\section{Conclusion}\label{sec:Conclusion}

In this work, we introduced \texttt{AegirJAX}, a fully differentiable non-hydrostatic hydrodynamic solver, to address the computational bottlenecks associated with inverse problems and physical parameterizations in coastal engineering. By formulating the fractional-step method and wet--dry boundary conditions to be entirely compatible with reverse-mode automatic differentiation, we demonstrated that a single computational framework can handle a diverse set of engineering tasks, comprising of forward, inverse and control problems. The differentiable paradigm exhibits profound strengths. Without the need to manually derive adjoint equations, we successfully performed continuous topology optimization for breakwater design and inverted complex, dynamic sources such as accelerating submarine landslides strictly from downstream gauge data. Furthermore, integrating neural networks directly into the physics loop allowed for the discovery of active, recurrent control policies for wave cancellation, and enabled the \texttt{AegirJAX} hybrid solver to dynamically correct truncation errors and missing dispersive physics in classical benchmarks like the Beji--Battjes, Monai Valley and Conical Island datasets.

Despite these successes, our experiments also highlight fundamental limitations inherent to data-driven physical modeling. As observed in the Beji--Battjes and Monai Valley dispersive correction experiments, the learned neural forcing exhibited strong regime-specific fidelity. Because the network was trained on sparse real-world experimental data, it effectively learned a local correction tailored to the specific wave kinematics of that dataset, lacking the diverse exposure required to generalize across vastly different wave periods and amplitudes. Historically, physical benchmarks such as the Conical Island and Monai Valley experiments were primarily designed for the point-validation of forward models, often relying on isolated or specific wave events. However, the emergence of scientific machine learning and its demonstrated ability to extract unresolved physics directly from sensor observations fundamentally broadens the utility of physical scale models. This paradigm strongly motivates a renewed investment in large-scale laboratory experiments. By systematically varying incident wave parameters across complex bathymetries, the experimental community could generate the rich, densely sampled datasets that are now necessary to train universally generalizable sub-grid parameterizations. Similarly, as shown in the bathymetry inversion task, regions devoid of downstream sensor influence remain mathematically unconstrained, reflecting the physical reality of the ill-posed inverse problem. Future research must address these generalization bottlenecks. A further promising avenue might be to pre-train sub-grid neural corrections on massive, synthetically generated datasets from high-fidelity three-dimensional Navier-Stokes simulations to establish a generalized physical baseline, before fine-tuning the models on sparse, real-world gauge data. 

Ultimately, we believe the fusion of differentiable physics and machine learning, as illustrated by \texttt{AegirJAX} for the particular problem of coastal engineering, represents a flexible and promising methodology for advancing the next generation of coastal hazard assessment and structural design algorithms that combines the best of both worlds of standard scientific computing and physics-informed machine learning.

\section*{Code and data availability statement}

The source code of \texttt{AegirJAX} is available for download on Github~\footnote{\url{https://github.com/abihlo/AegirJAX}}. The Beji--Battjes benchmark data reported in was generously provided by Professor Dr.\ Serdar Beji. The data for the Conical Island and Monai Valley benchmarks are described in~\cite{brig95a,liu95a} and~\cite{mats01a}, respectively. 

\section*{Statement on the use of generative artificial intelligence}

Gemini 3 Pro was used as coding assistant for developing the differentiable solver \texttt{AegirJAX} used in this paper.

\begin{ack}
This research was carried out, in part, through funding from the NSERC Discovery Grant program.
\end{ack}

{\footnotesize\setlength{\itemsep}{0ex}

}

\appendix

\section{Implementation and training details}\label{sec:Appendix}

All experiments were implemented using the \texttt{AegirJAX} solver, written natively in Python utilizing the \texttt{JAX} library for reverse-mode automatic differentiation and XLA compilation. Neural network architectures were constructed using the \texttt{flax} library, and optimizations were performed using the \texttt{optax} library. All spatial boundary conditions and wet--dry masking were formulated as fully differentiable, continuous array operations to avoid control-flow graph severing. Optimization was performed using the Adam optimizer coupled with either a fixed learning rate or with a cosine decay learning rate schedule. A global gradient norm clipping set to 1.0 to prevent unreasonably large gradient descent steps. Simulations were carried out on a single NVIDIA RTX 4090 GPU.

\subsection{Model discovery and physical correction}
 
For the Beji--Battjes, Monai Valley, and Conical Island benchmarks, the neural corrector utilized a dilated residual convolutional neural network with skip connections to ensure stable identity mappings. The network consisted of an initial projection layer followed by three hidden residual blocks with 32 channels and progressively increasing dilations of 1, 2, and 4 to expand the receptive field without pooling. Activations utilized the GELU function.

To guarantee exactly zero neural correction in quiescent states, the final projection layer omitted biases and was bounded by a scaled tangent hyperbolic activation, $0.1 \tanh(x)$. To enforce conservative forcing and prevent artificial energy injection, the network predicted a single-channel pseudo-pressure potential, which was then spatially differentiated using a second-order central difference stencil to yield the applied acceleration field. Furthermore, all physical inputs to the network were modulated by a localized input gate, applying a smooth sigmoid function over the local kinematic energy to mask out acoustic numerical noise in still water.

For the one-dimensional Beji--Battjes benchmark and the two-dimensional Monai Valley and Conical Island benchmarks, spatial Total Variation (TV) penalties with weights of $\lambda_{\rm TV} = 10^{-4}$ and $\lambda_{\rm TV} = 0.005$, respectively, were applied to the derived acceleration fields to suppress high-frequency numerical ringing.

\subsection{Inverse design and topology optimization}

The computational domain for both benchmarks was $\mathcal{D} = [0, 200] \times [0, 80]$, using a $200 \times 80$ grid with a spatial resolution of $\Delta x = \Delta y = 1.0$ m 

For the rigid-body breakwater gate positioning, an unperturbed water depth of $H_0 = 4.0$ m was used.  The simulation was integrated over 4,000 temporal steps using a time-step of $\Delta t = 0.005$ s. The static geometry featured a continuous wall at $x = 100$ m with a 20 m wide gap. The trainable parameters were defined as the continuous center coordinates $(x_c, y_c)$ and the rotation angle $\varphi$ of the rectangular gate. To ensure differentiability, the rigid gate was mapped onto the Eulerian grid utilizing a signed distance field and a continuous sigmoid rasterization function, $M(x,y) = \sigma(-k \cdot d)$, where~$k=5$ controlled the sharpness of the structural boundary. The harbor zone was defined as the region beginning 10 m downstream of the static sea wall, located at $x = 100$ m. Therefore, the energy minimization objective is evaluated over the domain $\Omega_{\rm harbor} = \{ (x,y) \in \mathcal{D} \mid x \geqslant 110 \}$, and the loss function was evaluated exclusively within the designated harbor zone by summing the local kinetic energy over the entire temporal rollout. 

For the construction of the optimal breakwater with a fixed material volume, integration was carried out over 10,000 temporal steps with a time-step of $\Delta t = 0.0025$ s. The latent design variables, i.e.\ the unconstrained logits, were mapped to physical density $\rho \in [0,1]$ via a sigmoid function and smoothed via a two-dimensional Gaussian convolution to encourage contiguous structures. The added breakwater height was defined via a cubic penalty, $\Delta h = \rho^3 H_{\rm max}$, with $H_{\rm max} = 2.5$ m. The total material volume was constrained to $V_{\rm target} = 500.0$ m$^3$, enforced via an absolute percentage error penalty in the loss function. The objective function seeks to protect a specific nearshore facility located on the sloping beach, just upstream of the final protected boundary. The wave impact penalty is evaluated exclusively within this localized patch, defined as $\Omega_{\rm facility} = \{ (x,y) \in \mathcal{D} \mid x \in [140, 160], \; y \in [30, 50] \}$.

\subsection{Active control of hydrodynamic systems} 

The one-dimensional wave cancellation policy was parameterized using a Gated Recurrent Unit (GRU). The computational domain spanned $L = 100.0$ m with a uniform static water depth of $H_0 = 2.0$ m, discretized using $\Delta x = 0.5$ m and integrated with a time step of $\Delta t = 0.02$ s over a total simulation time of 20.0 s (1000 temporal steps). A sensor array is located between $x=20$ m and $x=50$ m, an active wave-maker (actuator) is centered at $x=50$ m, and a protected harbor zone is defined for $x>75$ m.

The base model utilized the full non-hydrostatic predictor--corrector scheme. At each time step, the policy ingested the free-surface elevation and horizontal velocity from the sensor array (spanning $x \in [20, 50]$ m) into a GRU cell with a hidden state dimension of 32. The updated memory state was passed through a single dense layer with 32 units using the ReLU activation to output a scalar force command. The physical actuator force was bounded using $F_{\rm actuator} = 2.0 \tanh(x)$ to prevent violent, non-physical control actions. This force was applied to the fluid momentum via a spatial Gaussian influence profile, defined as $\exp(-(x - 50)^2 / 4.0)$. To ensure generalized learning, the model was trained on a batch of three solitary waves with varying amplitudes ($A \in \{0.15, 0.2, 0.35\}$ m) and initial release positions ($x_0 \in \{15.0, 20.0, 25.0\}$ m). The loss function utilized an actuator effort penalty weight of $\lambda_{\rm effort} = 0.01$. The policy was trained over 800 epochs using back-propagation through time across the fully unrolled fluid sequence.

\subsection{Source inversion and parameter estimation}

For the fixed bottom topography inversion, the simulation domain spanned $L=30.0$ m with $\Delta x = 0.05$ m, $\Delta t = 0.005$ s, and a rollout of $20.0$ s. The bathymetry was parameterized by a multi-layer perceptron with a hidden dimension of 128 utilizing Fourier feature embeddings using a frequency scale of 6.0, mapping the normalized spatial coordinate to the static depth $h$. The loss function was regularized using the $L_2$-norm of the spatial gradient to enforce smooth profiles, alongside a quadratic volume penalty to anchor the unperturbed tank floor at its known depth of $0.4$ m. Optimization was performed across a three-phase scheduling strategy: 100 epochs of heavy smoothing ($\lambda_{\rm smooth}=1.0, \lambda_{\rm vol}=10^{-4}$), 100 epochs of relaxed smoothing ($\lambda_{\rm smooth}=0.05, \lambda_{\rm vol}=10^{-4}$), and 1000 epochs of fine polishing ($\lambda_{\rm smooth}=10^{-5}, \lambda_{\rm vol}=10^{-5}$).

For the landslide inversion, the domain spanned $L=200.0$ m using a spatial grid of $\Delta x = 0.5$ m over a static sloped bed linearly increasing in depth from $1.0$ m to $15.0$ m. The time stepof the integration was $\Delta t = 0.015$ s.  The landslide was initialized at $x=30.0$ m with an initial acceleration of $0.8$ m/s$^2$ and a terminal velocity of $6.0$ m/s. The dynamic spatial footprint was parameterized by a base semi-elliptical shape and an additive multi-layer perceptron correction. Both the kinematic width parameter $b$ and the base thickness parameter $T$ were optimized jointly alongside the network weights. To enforce positivity, these scalars were parameterized in log-space (e.g., $b=\exp(\log b)$). Optimization was performed against two gauges located at $x \approx 100$ m and $150$ m, regularized by an $L_2$-penalty on the neural correction magnitude with a weight of $10^{-4}$ to prevent noise injection.

For the passive tracer source localization experiment, the computational domain was based on the Monai Valley bathymetry, downsampled by a stride of 4 to accelerate training. The simulation was integrated over a total temporal rollout of 28.0 s. The base hydrostatic solver was augmented with a depth-integrated advection-diffusion kernel, utilizing an upwind scheme for the advective fluxes and a central-difference approximation for the diffusion term. The diffusion coefficient was set to $K=0.1$ $\rm m^2/s$. The trainable parameters for the inversion were the continuous spatial coordinates of the spill, $x_c$ and $y_c$, and the unconstrained raw injection mass. To ensure the gradients could flow smoothly from the Eulerian grid to the continuous coordinate parameters, the point source was rasterized onto the computational grid using a differentiable spatial Gaussian with a standard deviation equal to twice the grid resolution. This spatial Gaussian was normalized such that its discrete spatial sum equaled one. The scalar mass parameter $M$ was then multiplied by this normalized distribution to form the continuous source term $S(x,y,t)$ in the advection--diffusion equation. Consequently, $M$ represents the total rate of tracer concentration injected into the entire domain per second. For this experiment, the synthetic ground-truth mass injection rate was set to $M = 5.0$. To enforce a strictly positive physical mass during the optimization process, the raw mass parameter was transformed using a softplus activation function. The objective function was formulated as the mean squared error between the simulated tracer concentrations and the ground-truth target concentrations recorded at the three downstream gauge locations. To prevent vanishing gradients from the initially small concentration values, the mean squared error loss was scaled by a factor of 1000.0.

\end{document}